# Pressure-stabilized dual-BCC polymorphism in a rhenium-based high-entropy alloy

Raimundas Sereika[*], Andrew D. Pope, Hunter Kantelis, Caleb M. Knight, Kallol Chakrabarty, and Yogesh K. Vohra

*Department of Physics, University of Alabama at Birmingham, Birmingham, Alabama 35294, USA*
*Corresponding author: rsereika@uab.edu*

**Abstract**

Accessing metastable structural states in high-entropy alloys offers a promising route to tailor material properties, yet the use of high pressure to engineer such states remains underexplored. Here, we report the pressure-driven synthesis of a unique metastable dual-BCC microstructure in a near-equimolar ReNbTiZrHf alloy. Starting from an ambient two-phase mixture of hexagonal (C14-derived) and body-centered cubic (BCC) phases, compression induces a selective, diffusionless transformation of the hexagonal constituent into a second, crystallographically distinct BCC polymorph, while the original BCC phase remains stable. Upon decompression, the pressure-induced BCC phase is kinetically trapped, yielding a dual-BCC state that is inaccessible via conventional thermal processing. The pressure-stabilized BCC polymorph is Re-enriched and inherits the exceptional stiffness of its hexagonal parent (bulk modulus ~290 GPa), creating a composite microstructure with pronounced elastic and mechanical contrast relative to the softer original BCC matrix (~180 GPa). These findings demonstrate that pressure can effectively navigate the flat free-energy landscapes of chemically complex alloys, establishing a robust pathway for polymorph engineering and metastable phase design in refractory HEAs.

# Introduction

High-entropy alloys (HEAs) represent a class of multicomponent metallic materials composed of five or more principal elements in near-equimolar ratios, where configurational entropy plays a key role in stabilizing simple crystal structures such as body-centered cubic (BCC), face-centered cubic (FCC), and hexagonal close-packed (HCP) lattices [1–3]. Despite their chemical complexity, many HEAs crystallize as single-phase solid solutions, combining structural simplicity with pronounced chemical disorder. This distinctive combination enables broad tunability of mechanical and physical properties, including high strength, hardness, corrosion resistance, thermal stability, and, in selected systems, functional electronic behavior such as superconductivity [2–5].

Rhenium-containing HEAs form a particularly intriguing subclass within this family. Rhenium itself crystallizes in a hexagonal structure and exhibits exceptional mechanical strength and thermal stability [6–8]. When incorporated into multicomponent alloys, rhenium can strongly influence phase stability, electronic structure, and mechanical behavior [9–15]. In the Re–Nb–Ti–Zr–Hf system, Marik et al. reported superconductivity in a near-equimolar alloy with a BCC structure ($Re_{0.16}Nb_{0.21}Ti_{0.20}Zr_{0.20}Hf_{0.23}$) and a superconducting transition temperature of approximately 5.3 K [10], demonstrating that superconductivity can persist in a chemically disordered, entropy-stabilized lattice. Subsequently, a Re-rich composition $Re_{0.56}Nb_{0.11}Ti_{0.11}Zr_{0.11}Hf_{0.11}$ was shown to adopt a hexagonal structure while remaining superconducting, with a transition temperature of approximately 4.4 K [11]. In this system, the crystal structure is strongly governed by composition and valence electron concentration: increasing rhenium content favors hexagonal symmetry, whereas lower rhenium content stabilizes the BCC structure [11]. Consequently, compositions near the nominally equimolar limit fall within a broad compositional regime where hexagonal and BCC phases compete and may coexist, reflecting a gradual crossover rather than a sharp structural boundary between the two symmetries.

High pressure therefore provides an effective thermodynamic parameter to modify the free-energy balance between these competing structural states and to stabilize crystal structures that are inaccessible through composition or temperature alone. For the hexagonal Re-rich alloy $Re_{0.6}Nb_{0.1}Ti_{0.1}Zr_{0.1}Hf_{0.1}$, *in situ* synchrotron X-ray diffraction measurements revealed a first-order hexagonal-to-BCC transformation initiating near 44 GPa and accompanied by a significant volume collapse of about 6 % [14]. Remarkably, the pressure-induced BCC phase was retained upon decompression, demonstrating metastable phase trapping. The transformation proceeds via a diffusionless, martensitic-like mechanism without detectable elemental segregation, highlighting the unusual role of chemical disorder in controlling phase stability under compression. This behavior is especially striking given the extreme structural stability of elemental rhenium under compression. Furthermore, the same Re-rich alloy $Re_{0.6}Nb_{0.1}Ti_{0.1}Zr_{0.1}Hf_{0.1}$ was shown to transform from a disordered BCC solid solution into an ordered B2-type superstructure when appropriate pressure–temperature pathways are applied [15]. This latter observation underscores the delicate competition between configurational entropy and enthalpy in Re-based compositionally complex alloys.

These findings motivate investigation of the nominally equimolar ReNbTiZrHf composition, which lies within this hexagonal–BCC coexistence regime and already exhibits simultaneous presence of both phases at ambient conditions. Such intrinsic phase coexistence indicates an even more delicate free-energy balance governed by competing enthalpic ordering tendencies and configurational entropy than in the Re-rich limit and raises fundamental questions regarding how pressure influences the evolution of this two-phase microstructure. Accordingly, in this work we investigate the high-pressure structural evolution of nominally equimolar ReNbTiZrHf to determine whether compression drives the alloy toward a single BCC phase, stabilizes distinct BCC variants, or promotes chemical ordering. The results extend the understanding of phase stability and metastability in Re-based compositionally complex alloys and highlight high pressure as an effective route for tuning structural, and potentially functional, complexity in entropy-stabilized materials.

## Methods

**Fabrication of the alloy**

Nominally equimolar ReNbTiZrHf alloy was synthesized using a MAM-1 vacuum arc melting system (Edmund Bühler GmbH, Bodelshausen, Germany). High-purity elemental metals (≥99.9%) were weighed according to the target stoichiometric composition and cold-pressed into a dense pellet using a hydraulic press. The pellet was placed on a water-cooled copper hearth inside the arc-melting chamber. Prior to melting, the chamber was evacuated and backfilled with high-purity argon to establish an inert atmosphere. The alloy was melted using a tungsten electrode. To ensure chemical homogeneity and microstructural uniformity, the solidified button was flipped and remelted four times, with each surface exposed to the arc. After synthesis, the bulk alloy was sectioned into smaller specimens using a diamond saw for subsequent characterization.

**Microstructural and Compositional Characterization**

All scanning electron microscopy and energy-dispersive X-ray spectroscopy (SEM–EDS) measurements were performed at room temperature and ambient pressure. Microstructural observations and quantitative point analyses were carried out using a Quanta FEG 650 scanning electron microscope (FEI, Hillsboro, OR, USA) operated at an accelerating voltage of 20 kV. Elemental mapping was performed using a JEOL JSM-7200F field-emission scanning electron microscope (JEOL USA, Peabody, MA, USA). The acquired EDS data were used to evaluate elemental distribution and phase partitioning within the alloy. Typical uncertainties of SEM–EDS quantitative analysis for individual elements are on the order of ±0.5–1 at.% and may vary depending on local composition, surface topography, and matrix effects.

**Synchrotron X-ray diffraction measurements and data processing**

High-pressure synchrotron X-ray diffraction (XRD) experiments were conducted using symmetric diamond anvil cells (DACs) equipped with standard diamonds of 250 μm culet size (run 1) and 100 μm culet size (run 2). In run 1, a rhenium gasket was precompressed to a thickness of 35 μm, and a 120 μm diameter hole was drilled to serve as the sample chamber. In run 2, the rhenium gasket was precompressed to 25 μm, and a 40 μm diameter hole was drilled. In both runs, powdered ReNbTiZrHf together with a copper pressure marker were loaded into the gasket hole. No pressure-transmitting medium was used. Pressure was determined from the copper using the equation of state reported by Sakai et al. [16] Taking into account the calibration uncertainty of the equation of state, fitting errors, and possible pressure gradients in the sample chamber, the absolute pressure uncertainty is estimated to be approximately 2–3% over the entire compression range (about ±1–2 GPa at 76 GPa and ±3–6 GPa at 169 GPa).

Angle-dispersive XRD measurements were performed at beamline 16-ID-B of the Advanced Photon Source (APS), Argonne National Laboratory (ANL). The incident monochromatic X-ray beam (energy 29.2 keV, wavelength 0.4246 Å) was focused to a 1.5 μm × 1.5 μm spot using Kirkpatrick–Baez mirrors. The wavelength was calibrated using a $CeO_2$ standard. Two-dimensional diffraction images were collected using a Pilatus 2M CdTe detector and azimuthally integrated into one-dimensional diffraction patterns with the DIOPTAS [17] software. Phase identification and structural refinements were carried out using Jana2020 [18]. Le Bail profile fitting was employed to determine lattice parameters and track the evolution of coexisting phases with pressure. Quantitative phase fraction analysis was not performed due to peak overlap between coexisting phases. Pressure–volume data were fitted using a third-order Birch–Murnaghan equation of state implemented in the EoSFit [19] software.

# Results

Scanning electron microscopy reveals that the nominally equimolar ReNbTiZrHf alloy exhibits a clearly resolved two-phase microstructure at ambient conditions (Fig. 1). Backscattered-electron SEM images show bright island-like regions embedded within a darker, continuous matrix. The bright islands are assigned to a hexagonal phase, while the darker matrix corresponds to a BCC phase (as explained below). Here, the energy-dispersive X-ray spectroscopy demonstrates pronounced chemical partitioning between the two phases. The BCC matrix is depleted in rhenium and enriched in Nb, Ti, and Zr, with an average composition of approximately $Re_{0.10}Nb_{0.24}Ti_{0.26}Zr_{0.22}Hf_{0.18}$. In contrast, the hexagonal island regions are strongly enriched in rhenium, with an average composition of approximately $Re_{0.39}Nb_{0.15}Ti_{0.16}Zr_{0.13}Hf_{0.17}$. Hafnium remains comparatively uniform between the two phases. The overall alloy composition averaged over the full scanned area is $Re_{0.20}Nb_{0.20}Ti_{0.23}Zr_{0.20}Hf_{0.17}$, confirming that the bulk material is close to, but not strictly, equimolar.

The corresponding EDS elemental maps (Fig. 2) qualitatively corroborate this phase partitioning. Among all elements, rhenium exhibits the strongest spatial contrast, with clear enrichment in the island regions, whereas Nb, Ti, and Zr are preferentially concentrated in the matrix phase. These observations demonstrate that the alloy separates into two chemically distinct solid-solution phases rather than forming a chemically homogeneous single-phase structure. The present SEM–EDS results therefore establish that near-equimolar ReNbTiZrHf lies within a compositional regime where hexagonal and BCC phases coexist, reflecting a delicate balance between competing structural preferences. This intrinsic two-phase state provides a unique starting point for investigating pressure-induced structural evolution in this alloy system.

At ambient pressure, the hexagonal phase has been well characterized in Re-rich compositions and was identified as a 12-atom C14 Laves phase with the $P6_3/mmc$ space group (No. 194), with lattice parameters $a \approx 5.257$ Å and $c \approx 8.615$ Å for $Re_{0.6}Nb_{0.1}Ti_{0.1}Zr_{0.1}Hf_{0.1}$ [14,15] and $a \approx 5.255$ Å and $c \approx 8.593$ Å for $Re_{0.56}Nb_{0.11}Ti_{0.11}Zr_{0.11}Hf_{0.11}$ [11]. In contrast, the BCC phase stabilized in Re-poor compositions corresponds to a disordered solid solution in the $Im\bar{3}m$ space group (No. 229), with a lattice parameter approximately $a$ = 3.380 Å for $Re_{0.16}Nb_{0.21}Ti_{0.20}Zr_{0.20}Hf_{0.23}$ [10]. Consistent with these reports, our synchrotron X-ray diffraction (XRD) measurements yield closely comparable lattice parameters for the present $Re_{0.20}Nb_{0.20}Ti_{0.23}Zr_{0.20}Hf_{0.17}$ alloy, with $a \approx 5.360$ Å and $c \approx 8.641$ Å for the hexagonal phase and $a \approx 3.400$ Å for the BCC phase.

Figure 3 summarizes the high-pressure structural evolution of nominally equimolar $Re_{0.20}Nb_{0.20}Ti_{0.23}Zr_{0.20}Hf_{0.17}$ obtained by in situ synchrotron X-ray diffraction in a diamond anvil cell. Representative diffraction patterns collected before compression and after completion of a full compression–decompression cycle are shown in Figs. 3a and 3b, respectively, and the corresponding pressure–volume relations are compiled in Fig. 3c. At ambient pressure, the diffraction pattern confirms coexistence of a hexagonal phase and a BCC phase, hereafter denoted BCC1 (Fig. 3a), consistent with the two-phase microstructure identified by SEM–EDS. Upon compression, reflections from the hexagonal phase progressively decrease in intensity and are replaced by a second set of BCC reflections (denoted BCC2). The emergence and growth of BCC2 at the expense of the hexagonal peaks indicate a pressure-induced hexagonal-to-BCC transformation of the hexagonal component, while the pre-existing BCC1 phase remains present and simply shifts continuously with pressure without evidence of an additional structural change. This transformation initiates near the mid-40 GPa range and proceeds over an extended pressure interval of roughly 30 GPa, as reflected by the prolonged coexistence of hexagonal peaks with BCC2 and by the corresponding evolution in the pressure–volume data (Fig. 3c). An extended two-phase interval is not unexpected for chemically complex alloys compressed at room temperature. In the Re-rich analogue $Re_{0.6}Nb_{0.1}Ti_{0.1}Zr_{0.1}Hf_{0.1}$, the hexagonal-to-BCC transition likewise exhibits a pressure range where only the parent hexagonal and product BCC phases coexist, with reproducibility across runs indicating that chemical disorder is a primary contributor, while minor pressure gradients may further broaden the apparent coexistence window in DAC experiments [14].

After decompression from 76 GPa to ambient pressure, the diffraction pattern shows retention of two distinct BCC phases (Fig. 3b). The phases are readily distinguished by their lattice parameters, with BCC1 refining to $a \approx 3.400$ Å and BCC2 to $a \approx 3.255$ Å. Notably, hexagonal reflections are not recovered after pressure release. Thus, the nominally equimolar alloy evolves from an initial hexagonal + BCC state into a metastably retained dual-BCC state after a single pressure cycle. The inset of Fig. 3c shows the pressure dependence of the hexagonal $c/a$ ratio prior to transformation. The $c/a$ ratio remains nearly constant over the entire stability range of the hexagonal phase, indicating that the lattice compresses in an almost isotropic manner before the onset of the structural instability. In combination, these results establish that in run 1 compression selectively converts the hexagonal constituent into a second BCC variant that is quenchable to ambient conditions alongside the original BCC phase, providing the first indication of pressure-stabilized dual-BCC metastability in near-equimolar Re–Nb–Ti–Zr–Hf.

Figure 4 shows the high-pressure synchrotron X-ray diffraction results obtained during the second compression–decompression cycle using smaller culet diamonds, allowing pressures up to approximately 169 GPa. The diffraction patterns collected during compression (Fig. 4a) and decompression (Fig. 4b), together with representative patterns at the highest pressure and after full pressure release (Figs. 4c and 4d), confirm the reproducibility of the transformation sequence observed in run 1. At low pressures, the alloy again exhibits coexistence of a hexagonal phase and a BCC phase (BCC1). Upon compression, the hexagonal reflections progressively weaken and are replaced by reflections of a second BCC phase (BCC2), indicating a pressure-induced hexagonal-to-BCC transformation of the hexagonal component. No additional structural transitions are detected up to the maximum pressure of about 169 GPa, demonstrating that the transformation pathway remains unchanged over the extended pressure range.

At the highest pressures, diffraction peaks from BCC1 and BCC2 approach each other and nearly merge within experimental resolution (Fig. 4c), reflecting the convergence of their lattice parameters under extreme compression. Here the refined lattice parameters are $a \approx 2.908$ Å for BCC1 and $a \approx 2.891$ Å for BCC2. However, upon decompression the two BCC phases separate into clearly distinguishable sets of reflections (Fig. 4d), confirming that the two cubic phases remain crystallographically distinct and are not fully homogenized at high pressure. The run-2 experiment thus independently confirms that near-equimolar ReNbTiZrHf undergoes a reproducible pressure-driven evolution from an initial hexagonal + BCC state into a metastable dual-BCC state, with no evidence of additional high-pressure phases up to 169 GPa.

Figure 5 compares the pressure–volume relations of the near-equimolar $Re_{0.20}Nb_{0.20}Ti_{0.23}Zr_{0.20}Hf_{0.17}$ alloy with those of the Re-rich $Re_{0.6}Nb_{0.1}Ti_{0.1}Zr_{0.1}Hf_{0.1}$ alloy, which exists as a single hexagonal phase at ambient conditions and whose high-pressure behavior has been reported previously (see Ref. 14). For the near-equimolar alloy, the hexagonal phase exhibits a pronounced volume discontinuity at the pressure-induced hexagonal-to-BCC transformation, with a volume collapse of approximately 6%, in close agreement with the behavior observed in the Re-rich alloy. The close similarity in collapse magnitude demonstrates that this volume reduction is largely insensitive to overall composition and instead represents an intrinsic characteristic of the hexagonal-to-BCC transformation in Re-based compositionally complex alloys. At comparable pressures, however, the Re-rich alloy consistently displays smaller atomic volumes than the near-equimolar alloy, indicating a systematically higher density. This trend is consistent with the higher rhenium content and confirms that rhenium plays a dominant role in governing both the density and compressibility of these alloys under extreme compression.

Fitting the pressure–volume data of $Re_{0.6}Nb_{0.1}Ti_{0.1}Zr_{0.1}Hf_{0.1}$ using a third-order Birch–Murnaghan equation of state yields bulk moduli of $K_0 = 279 \pm 2$ GPa with $K'_0 = 5.9$ for the hexagonal phase and $K_0 = 286 \pm 5$ GPa with $K'_0 = 4.0$ for the pressure-induced BCC2 phase. These high bulk moduli reflect the exceptional incompressibility of both phases and are comparable to those of refractory transition-metal alloys. For the near-equimolar $Re_{0.20}Nb_{0.20}Ti_{0.23}Zr_{0.20}Hf_{0.17}$ alloy, the pre-existing BCC1 phase exhibits a significantly lower bulk modulus, approximately in the range of 174–180 GPa within fitting uncertainty. In contrast, both the hexagonal phase and the pressure-induced BCC2 phase display substantially higher stiffness, with bulk moduli approaching 280–290 GPa. This contrast indicates that the pressure-induced BCC2 phase inherits the high incompressibility of the hexagonal parent phase, whereas the original BCC1 phase retains a more compressible solid-solution character. It should be noted that the bulk modulus values obtained in the present work differ slightly from those reported in Ref. [14], where the equation-of-state analysis was limited to a lower pressure range (up to ~70 GPa). In contrast, the present measurements extend to ~170 GPa, providing a broader fitting range and a more robust determination of the elastic parameters. Despite these differences, both studies consistently demonstrate the exceptionally high incompressibility of the Re-rich hexagonal and pressure-induced BCC2 phases.

## Discussion

From a thermodynamic perspective, the present results indicate that different stabilization mechanisms govern the three phases involved. At ambient conditions, the Re-rich hexagonal phase is likely stabilized predominantly by enthalpic and topological factors associated with Re bonding and C14-type packing, whereas BCC1 is consistent with an entropy-stabilized, chemically disordered solid solution, as commonly observed in refractory HEAs with reduced valence electron concentration and weaker directional bonding constraints. Under compression, the relevant competition is therefore not between the hexagonal phase and BCC1, but between the hexagonal phase and the pressure-induced BCC2 phase that emerges from it. This behavior reflects the flat free-energy landscape characteristic of high-entropy alloys, in which competing crystal structures are separated by small energetic differences, allowing pressure to act as an efficient selector of metastable polymorphs.

No evidence of B2 or any other long-range chemical order was detected during room-temperature compression up to 169 GPa. All diffraction peaks associated with BCC2 can be indexed by a disordered BCC lattice, with no superlattice reflections observed. This indicates that, at 300 K, atomic mobility is insufficient to enable long-range ordering during or after the transformation. Pressure alone therefore drives a structural transition but does not overcome kinetic barriers to chemical ordering. The contrast with combined pressure–temperature experiments, where ordered superstructures can be realized in related Re-rich compositions, highlights the essential role of thermal activation in accessing additional configurational states [15].

The hexagonal → BCC transformation exhibits clear signatures of a diffusionless, martensitic-like mechanism. The progressive disappearance of hexagonal reflections and the concurrent growth of the BCC2 phase over a broad pressure interval of approximately 30 GPa indicate a sluggish, first-order transition with extended coexistence interval. Such behavior is consistent with a transformation governed by cooperative lattice shear and shuffle rather than by long-range atomic diffusion. The conversion is therefore best described as a coordinated distortion of the C14-derived hexagonal framework into the cubic BCC2 lattice, while preserving the local chemical heterogeneity of the alloy.

Comparable sluggish, martensitic pressure-induced transformations have been reported in other compositionally complex alloys. In the equiatomic CrMnFeCoNi (Cantor) alloy, a pressure-driven FCC → HCP transition proceeds over more than 40 GPa and yields a metastable two-phase state upon decompression [20]. The extended coexistence interval observed in the present alloy, with hexagonal and BCC2 phases persisting over tens of gigapascals, can likewise be attributed to chemical disorder and local stress variations that distribute the transformation barrier across the microstructure [2,21]. Similar transition broadening has been widely recognized as a hallmark of phase transformations in chemically complex alloys [21]. The observation that diffusionless polymorphic transitions also occur in simpler systems, such as the HCP → BCC transformation in titanium under pressure [22], further supports a martensitic interpretation for the present hexagonal → BCC2 transformation.

Applying high pressure introduces a mechanical work contribution to the Gibbs free energy. In simplified form, the pressure dependence of the free-energy difference between the hexagonal phase and BCC2 may be written as:

$$\Delta G(P) \approx \Delta G(0) + \int_0^P \Delta V dP \approx \Delta H_{\text{mix}} - T\Delta S_{\text{config}} + P\Delta V,$$

where $\Delta V$ is the specific volume difference between phases. In this alloy, the hexagonal → BCC2 transformation is accompanied by an approximately 6% volume collapse. Consequently, increasing pressure progressively lowers the Gibbs energy of BCC2 relative to the hexagonal phase. At the observed transformation onset (around 45 GPa), the $P\Delta V$ term associated with the hexagonal → BCC2 volume collapse ($\Delta V \approx 0.9$ Å$^3$ per atom) corresponds to approximately 0.25 eV per atom ($\approx 25$ kJ mol$^{-1}$), providing a substantial driving force favoring stabilization of the BCC2 phase relative to the enthalpically stabilized hexagonal structure. While this thermodynamic argument explains the transformation of the hexagonal phase into BCC2, it also raises the question of why the pre-existing BCC1 phase remains stable under compression. It is important to note that the stability of the pre-existing BCC1 phase cannot be explained solely by its lower bulk modulus. A lower bulk modulus corresponds to higher compressibility and may, in principle, allow a phase to reduce its Gibbs free energy more efficiently with increasing pressure through the $P\Delta V$ term. However, in the present system, the relevant thermodynamic competition is not between BCC1 and the hexagonal phase, but between the hexagonal phase and the pressure-induced BCC2 phase, which are structurally and chemically related. The BCC2 phase emerges from the hexagonal parent through a diffusionless transformation and retains its local chemical character inherited from the hexagonal parent, consistent with its high incompressibility. In contrast, BCC1 represents a chemically distinct solid solution that occupies a separate free-energy basin. Therefore, the persistence of BCC1 reflects not only its higher compressibility but also its distinct composition and the absence of a direct transformation pathway linking it to the hexagonal phase under the present conditions. Pressure thus drives the system across the hexagonal → BCC2 stability boundary, while BCC1 remains as an independent, coexisting metastable basin.

An important materials insight from this work is the role of rhenium partitioning in governing phase properties under compression. The equation-of-state analysis shows that BCC2 exhibits a remarkably high bulk modulus of approximately 280–290 GPa, comparable to that of the original hexagonal parent phase and consistent with the incompressible character of Re-rich frameworks [8,14]. In contrast, BCC1 is substantially more compressible, with a bulk modulus of roughly 170–180 GPa. This pronounced elastic contrast reflects chemical partitioning between the two cubic phases: BCC2 is inferred to be Re-enriched, whereas BCC1 is relatively Re-depleted.

The key implication is that pressure unifies crystallographic symmetry while preserving chemical identity, so that the high incompressibility of the hexagonal parent is retained as a form of chemical memory in BCC2 rather than averaging into the softer BCC1 matrix. In this sense, pressure provides a route to engineer mechanical heterogeneity within a single nominal chemical composition by converting one constituent into an ultra-stiff BCC polymorph while leaving the other as a more compliant BCC solid solution. As a result, the metastable dual-BCC microstructure combines a rigid, Re-enriched cubic phase with a more deformable BCC matrix, suggesting a potentially favorable balance of strength and ductility that could be further optimized by tuning pressure–temperature pathways. Future studies combining controlled pressure–temperature pathways with mechanical testing will be essential to determine whether the pressure-engineered dual-BCC microstructure can be translated into improved macroscopic mechanical performance.

## Conclusions

In summary, we report the pressure-driven synthesis of a unique metastable dual-BCC state in a nominally equimolar ReNbTiZrHf high-entropy alloy, originating from an ambient two-phase (hexagonal + BCC) microstructure. Compression induces a selective, martensitic-like transformation of the hexagonal constituent into a second, distinct BCC polymorph, while the original BCC matrix remains structurally intact. This transformation proceeds over a broad pressure interval and is accompanied by a characteristic volume collapse of approximately 6%, consistent with a lattice-instability-driven mechanism rather than diffusive rearrangement.

Upon decompression, the pressure-induced BCC polymorph is kinetically trapped, resulting in a reproducible metastable dual-BCC assemblage that is inaccessible via conventional thermal processing routes. The resulting microstructure exhibits pronounced mechanical heterogeneity, combining an ultra-stiff, Re-enriched BCC phase with a more compliant BCC solid-solution matrix.

These findings establish high pressure as a powerful pathway-engineering tool for accessing non-equilibrium structural states in compositionally complex alloys. By exploiting the interplay between configurational entropy, enthalpic bonding preferences, and mechanical work, this approach significantly expands the accessible phase space and property-design landscape of refractory high-entropy alloys.

**Acknowledgements**

This material is based upon work supported by the National Science Foundation (NSF) under Grant No. DMR-2310526. Portion of this work was performed at the HPCAT (Sector 16), Advanced Photon Source (APS), Argonne National Laboratory. HPCAT operations are supported by DOE-NNSA's Office of Experimental Sciences. The Advanced Photon Source is a U.S. Department of Energy (DOE) Office of Science User Facility operated for the DOE Office of Science by Argonne National Laboratory under Contract No. DE-AC02-06CH11357. The authors would like to thank Dr. Vijaya K. Rangari and Dr. Abhinav Yadav, Department of Materials Science and Engineering at Tuskegee University for assistance in Scanning Electron Microscopy and elemental mapping of high entropy alloys.

**Funding Declaration**

This research received funding from the NSF. Grant No. DMR-2310526

**Competing financial interests**

The authors declare that they have no competing financial interests.

**Data Availability**

The datasets used and analyzed during the current study are available from the corresponding author on reasonable request.

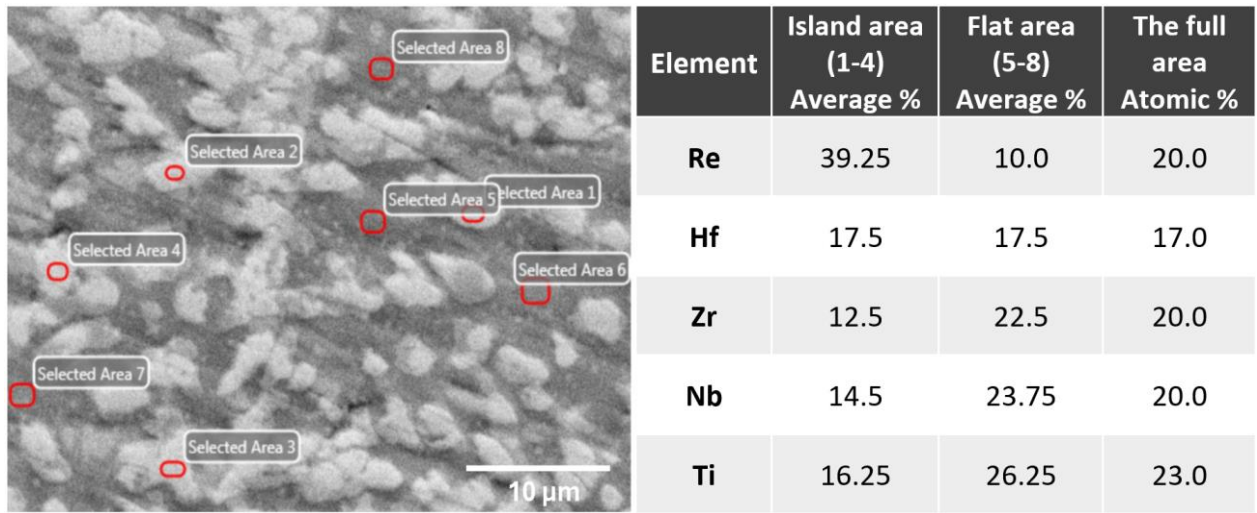

**Figure 1.** Backscattered-electron SEM image of the near-equimolar Re–Nb–Ti–Zr–Hf alloy showing a two-phase microstructure with bright island-like regions embedded in a darker matrix. Red markers indicate representative EDS point analyses. The table summarizes average atomic concentrations for island regions (areas 1–4), matrix regions (areas 5–8), and the full analyzed area, revealing strong partitioning of Re between the two microstructural constituents and corresponding variations in the other elements.

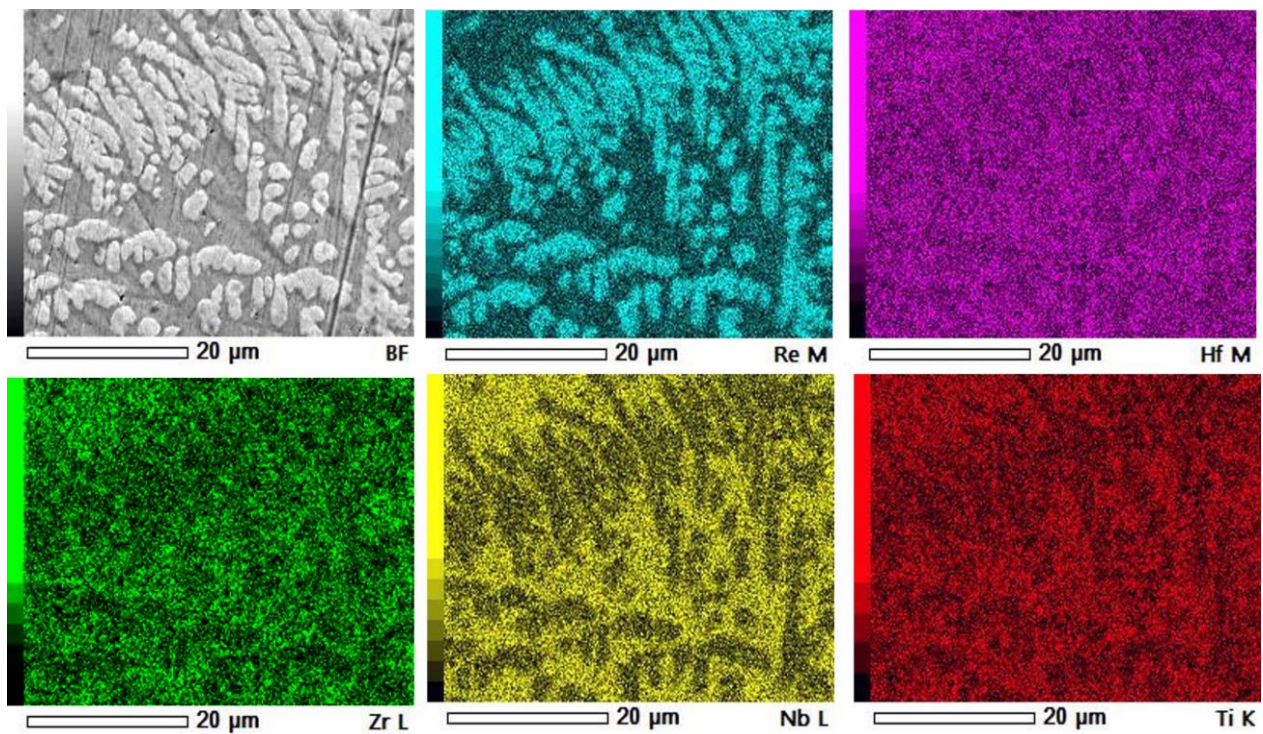

**Figure 2.** Backscattered-electron image (BF/BSE) and corresponding EDS elemental maps for Re, Hf, Zr, Nb, and Ti in the near-equimolar Re–Nb–Ti–Zr–Hf alloy. The maps qualitatively confirm compositional modulation correlated with the two-phase microstructure, with the strongest contrast observed for Re. (Elemental maps are displayed on independent intensity scales and are intended for qualitative comparison.)

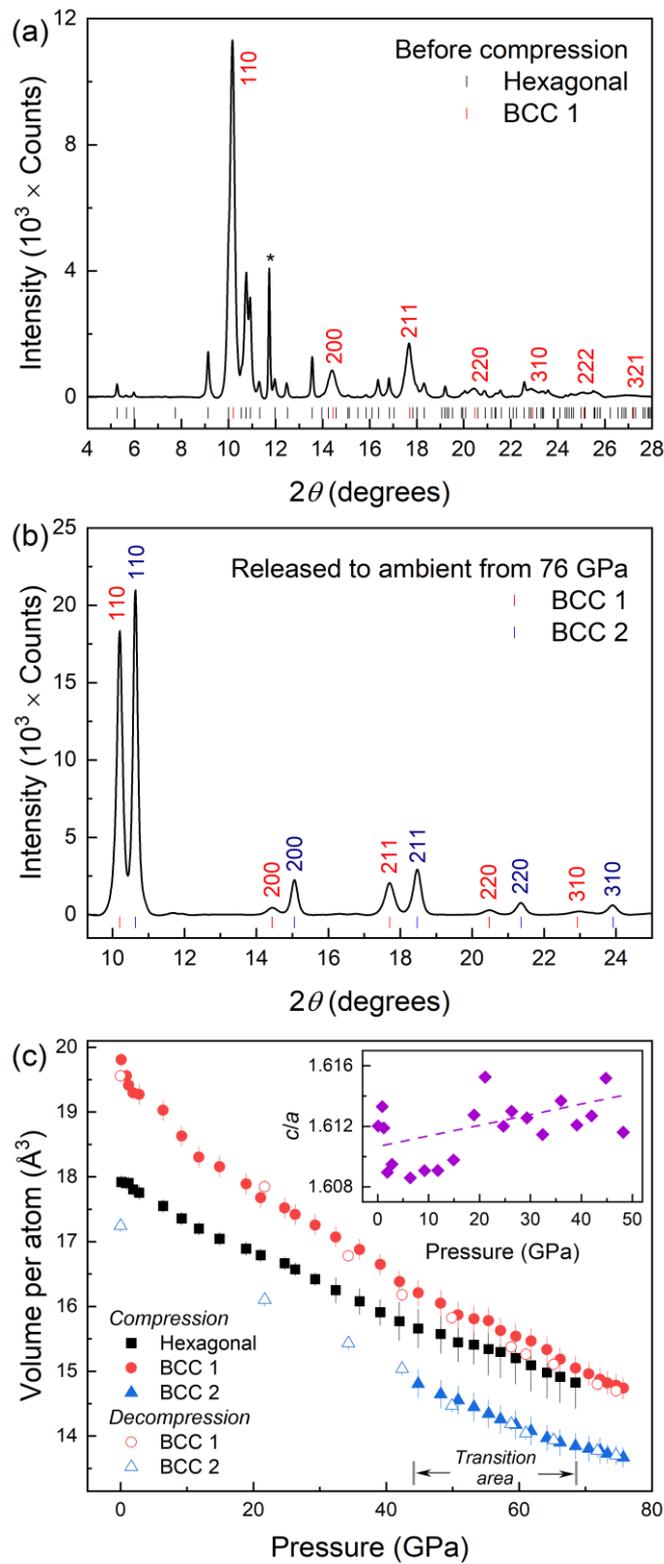

**Figure 3. High-pressure structural evolution of Re$_{0.2}$Nb$_{0.2}$Ti$_{0.23}$Zr$_{0.2}$Hf$_{0.17}$ from run 1. a**, Synchrotron X-ray diffraction pattern at ambient pressure showing coexistence of a hexagonal phase and a cubic phase (BCC1). Indexed Bragg peaks of BCC1 are labeled in red, while hexagonal reflections are indicated by black markers. The star denotes the Cu pressure marker. **b**, Diffraction pattern after decompression from 76 GPa, revealing two distinct cubic phases, denoted BCC1 (red indices) and BCC2 (blue indices). **c**, Pressure–volume relations for the hexagonal phase, BCC1, and BCC2 during compression and decompression. The inset shows the pressure dependence of the hexagonal $c/a$ ratio prior to transformation.

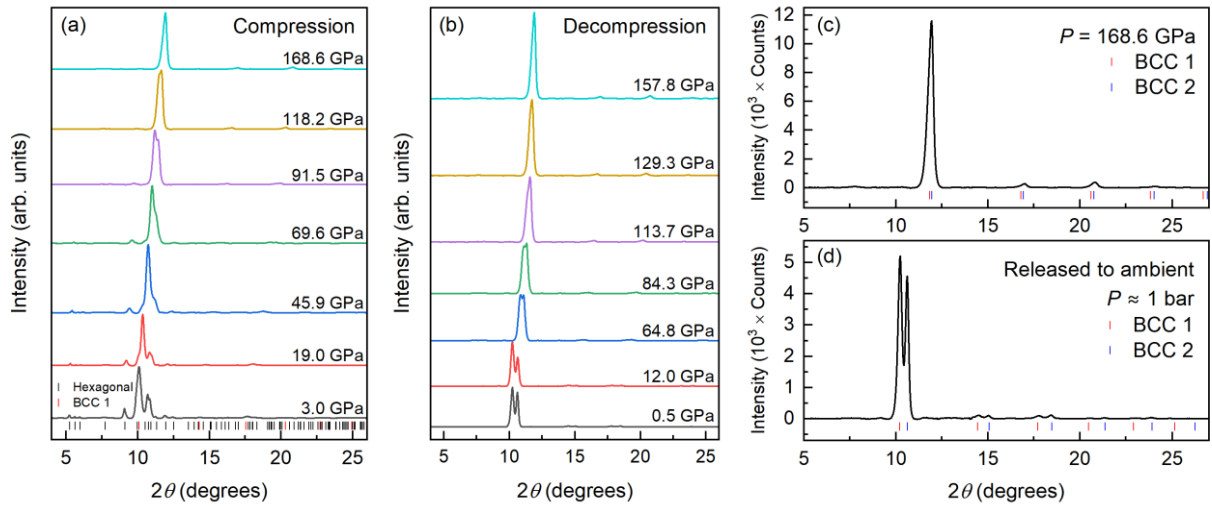

**Figure 4. High-pressure X-ray diffraction results for $Re_{0.20}Nb_{0.20}Ti_{0.23}Zr_{0.20}Hf_{0.17}$ from run 2. a**, Selected diffraction patterns collected during compression up to 168.6 GPa. **b**, Selected diffraction patterns collected during decompression. **c**, Diffraction pattern at the highest pressure (168.6 GPa), where reflections from BCC1 and BCC2 approach each other and nearly overlap. **d**, Diffraction pattern after full pressure release to ambient conditions, showing separation of BCC1 and BCC2 reflections and confirming retention of the dual-BCC state.

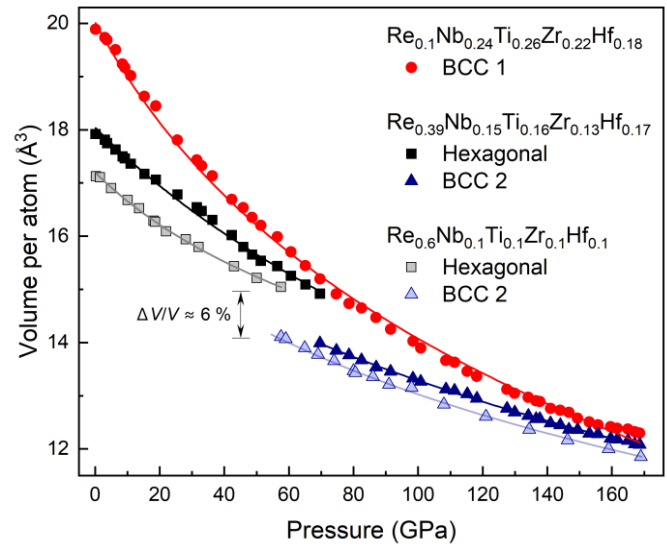

**Figure 5. Pressure–volume relations for the phase-separated constituents of the near-equimolar Re$_{0.20}$Nb$_{0.20}$Ti$_{0.23}$Zr$_{0.20}$Hf$_{0.17}$ alloy and the Re-rich Re$_{0.6}$Nb$_{0.1}$Ti$_{0.1}$Zr$_{0.1}$Hf$_{0.1}$ alloy.** For the near-equimolar alloy, data are shown separately for the BCC matrix phase (Re$_{0.1}$Nb$_{0.24}$Ti$_{0.26}$Zr$_{0.22}$Hf$_{0.18}$) and the hexagonal/BCC2 phase (Re$_{0.39}$Nb$_{0.15}$Ti$_{0.16}$Zr$_{0.13}$Hf$_{0.17}$). Solid lines represent third-order Birch–Murnaghan equation-of-state fits. A volume collapse of approximately 6% accompanies the hexagonal-to-BCC transformation in both alloys. At comparable pressures, the Re-rich alloy exhibits systematically smaller atomic volumes, indicating higher density associated with increased rhenium content.